\documentclass[twocolumn,showpacs,preprintnumbers,amsmath,amssymb]{revtex4}

\usepackage{graphicx}
\usepackage{dcolumn}
\usepackage{bm}

\begin{document}

\title{Spin effects in electron tunnelling through a quantum dot coupled
to non-collinearly polarized ferromagnetic leads}
\author{W.~Rudzi\'nski$^{1}$}
\author{J.~Barna\'s$^{1,2}$}
\author{R. \'Swirkowicz$^{3}$}
\author{M. Wilczy\'nski$^{3}$}
\affiliation{$^1$Department of Physics, Adam Mickiewicz
University, ul. Umultowska 85, 61-614 Pozna\'n, Poland \\
$^2$ Institute of Molecular Physics, Polish Academy of Sciences, \\
ul. Smoluchowskiego 17, 60-179 Pozna\'n, Poland \\
$^3$ Faculty of Physics, Warsaw University of Technology,  ul.
Koszykowa 75, 00-662 Warszawa, Poland}

\date{\today}

\begin{abstract}

Spin-dependent transport through an interacting single-level
quantum dot coupled to ferromagnetic leads with non-collinear
magnetizations is analyzed theoretically. The transport properties
and average spin of the dot are investigated within the
nonequilibrium Green function technique based on the equation of
motion in the Hartree-Fock approximation. Numerical results show
that Coulomb correlations on the dot and strong spin polarization
of the leads significantly enhance precession of the average dot
spin around the effective molecular field created by the external
electrodes. Moreover, they also show that spin precession may lead
to negative differential conductance in the voltage range between
the two relevant threshold voltages. Nonmonotonous angular
variation of electric current and change in sign of the tunnel
magnetoresistance are also found. It is also shown that the
diode-like behavior in asymmetrical junctions with one electrode
being half-metallic is significantly reduced in noncollinear
configurations.

\pacs{72.25.Mk; 73.63.Kv; 73.23.Hk}
\end{abstract}

\maketitle

\section{Introduction}

Current interest in electronic transport and spin effects in
mesoscopic tunnel junctions is stimulated by their possible
applications in microelectronics and spintronics devices
\cite{prinz99,daughton99}. One of the most widely studied
spin-dependent effects in magnetic tunnel junctions is the tunnel
magnetoresistance (TMR). This phenomenon appears as a change in
the junction resistance when magnetic moments of external
electrodes rotate from parallel alignment to non-collinear one (or
to antiparallel alignment in a particular case). Such a rotation
of magnetic moments may be induced, for instance, by an external
magnetic field. The TMR effect may occur in planar junctions
\cite{moodera95,parkin99,wilczynski00}, mesoscopic double-barrier
junctions \cite{barnas98,barnas99,takahashi98,brataas99}, granular
systems \cite{imamura00,yakushiji01}, and others. When the central
part of a double-barrier junction is sufficiently small, the
interplay of discrete charging by single electrons and spin
dependence of tunnelling processes can lead to additional
interesting features in the corresponding transport
characteristics \cite{barnas98,barnas99,takahashi98,brataas99}.

Up to now, most of theoretical works on transport through quantum
dots (QDs) coupled to ferromagnetic electrodes was limited to
collinear, i.e., parallel and antiparallel magnetic configurations
\cite{bulka00,rudzinski01,swirkowicz02,fransson02,bulka03,martinek03,lopez03}.
It is only very recently when transport in systems with
non-collinear magnetizations was addressed
\cite{sergueev02,barnas03,koenig03,braun04}. In particular, it has
been shown that the diode-like features in transport
characteristics of systems with one electrode being half-metallic
are significantly reduced when magnetic moments of the electrodes
become non-collinear \cite{barnas03}. However, this behavior was
studied only in the sequential tunneling regime.

In recent papers \cite{koenig03,braun04} spin precession in
electron tunnelling through an interacting quantum dot was studied
theoretically in the first order approximation with respect to the
tunneling Hamiltonian. Such a precession takes place when magnetic
moments of the leads are non-collinear, and occurs because an
electron entering the dot in a tunnelling event is subject to an
effective exchange field, which exerts a torque on the electron
spin and makes the spin precesses by a certain angle before the
electron leaves the dot. In the first order approximation, the
spin precession is driven by the Coulomb interaction on the dot
(described by the Hubbard correlation parameter $U$) and
disappears when $U$ tends to zero.

In this paper we extend the earlier descriptions
\cite{barnas03,koenig03,braun04} of electron tunneling through QD
with non-collinear magnetizations by going beyond the first order
approximation. In order to calculate the tunneling current and
spin precession, we employ the non-equilibrium Green function
technique and limit considerations to the Hartree-Fock
approximation. Hence, besides the sequential tunneling also the
contribution due to higher order tunneling processes is included
in the description. Numerical results show, that spin precession
also exist in the limit of $U=0$. However, Coulomb correlations
significantly enhance the precession. This, in turn, may lead to
negative differential conductance in a certain bias voltage range.
We predict that symmetry of the junction, barrier height, and spin
polarization of magnetic leads may significantly influence spin
dependent characteristics of the dot. In particular, it is shown
that the diode-like behavior in asymmetrical junctions with one
half-metallic electrode is partially suppressed in non-collinear
configurations. However, the suppression is much less evident than
that obtained within a simplified theory neglecting the exchange
interactions between the dot and leads \cite{barnas03}.

The paper is organized as follows. In Section II we describe
model of the system. Theoretical method is described in Section
III, where the equation of motion method is used to derive
nonequilibrium Green functions of the dot. Transport
characteristics are calculated in Section IV,  whereas relevant
numerical results on tunnelling current, magnetoresistance, and
spin precession are presented and discussed in Section V. Finally,
summary and general conclusions are in Section VI.

\section{Model}

We consider a single-level QD coupled to two ferromagnetic
metallic leads by tunnelling barriers. Magnetic moments in the
external leads lie in a common plane and form an arbitrary angle
$\varphi$. To describe electron spin we will use the local and
global quantization axes. The local axes are determined by the
local spin polarization in the leads. The global quantization axis
(the axis $z$ in our case) is assumed to coincide with the local
one in the left electrode. Spin projection on the local
quantization axes will be denoted as $\beta =+$ for majority
electrons and $\beta =-$ for minority electrons, whereas
projection on the global quantization axis will be denoted as
$\sigma =\uparrow$ and $\sigma =\downarrow$. Axis $y$ of the
coordinate system is normal to the plane $xz$ determined by the
spin polarizations of the leads. Geometry of the device and the
orientation of the coordinate system are shown schematically in
Fig.1.
\begin{figure}[h]
\begin{center}
\includegraphics[width=1\columnwidth,angle=0]{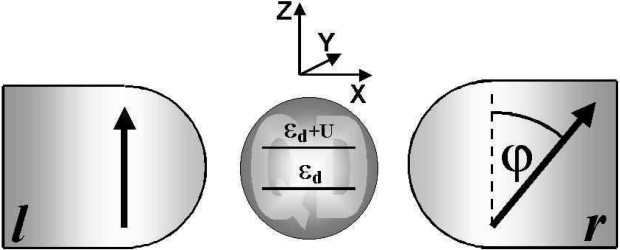}
\caption{\label{currentplots}Schematics of the system considered
in this paper. The coordinate systems used to describe states of
the dot is also shown.}
\end{center}
\end{figure}

The whole system can be described by Hamiltonian of the general
form
\begin{equation}
H=H_{l}+H_{r}+H_{d}+H_{t}.
\end{equation}
The term $H_{\nu}$ describes the left ($\nu =l$) and right ($\nu
=r$) electrodes in the non-interacting quasi-particle
approximation,
\begin{equation}
H_{\nu}=\sum_{k}\sum_{\beta =+,-}\varepsilon_{k\beta}^{\nu}a_{\nu
k\beta} ^{+}a_{\nu k\beta},
\end{equation}
where $\varepsilon_{k\beta}^{\nu}$ is the single-electron energy
in the $\nu$-th electrode for the wavevector k and spin $\beta$,
whereas $a_{\nu k\beta}^{+}$ and $a_{\nu k\beta}$ are the
corresponding creation and annihilation operators. The term
$H_{d}$ in Eq.(1) describes QD,
\begin{equation}
H_{d}=\sum_{\sigma}\varepsilon_{d}c_{\sigma}^{+}c_{\sigma}\
+Un_{\uparrow}n_{\downarrow},
\end{equation}
where $n_\sigma =c^+_\sigma c_\sigma$ is the occupation operator,
$\varepsilon_{d}$ denotes the energy of the discrete level, $U$ is
the electron correlation parameter, whereas $c_{\sigma}^{+}$ and
$c_{\sigma}$ are the corresponding creation and annihilation
operators for electrons with spin orientation $\sigma =\, \uparrow
(\,\downarrow)$. Finally, the tunnelling term, $H_t$, in Eq.(1)
takes the form
\begin{equation}
H_t=H_{t}^l+H_{t}^r ,
\end{equation}
where the first term describes tunnelling through the left
barrier,
\begin{equation}
H_{t}^l= \sum_{k}(T_{k+}^la^+_{lk+}c_\uparrow +
T_{k-}^la^+_{lk-}c_\downarrow )+ {\rm h.c.},
\end{equation}
whereas the second term corresponds to tunnelling through the
right barrier,
\[
H_{t}^r=\sum_{k}\{ [ T_{k+}^ra^+_{rk+}\cos (\varphi /2)-
T_{k-}^ra^+_{rk-}\sin (\varphi /2) ]c_\uparrow
\]
\begin{equation}
+ [T_{k-}^ra^+_{rk-}\cos (\varphi /2)+ T_{k+}^ra^+_{rk+}\sin
(\varphi /2)]c_\downarrow  \} +{\rm h.c.},
\end{equation}
and ${\rm h.c.}$ stands for the hermitian conjugate terms. The
tunneling terms $H_{t}^l$ and $H_{t}^r$ have different forms
because the corresponding local and global axes are parallel for
the left electrode and noncollinear for the right one.

\section{Green functions of the dot}

To calculate electric current in nonequilibrium situations we will
make use of the nonequilibrium Green function defined on the
Keldysh contour \cite{haug_book}. The causual Green function of
the dot is defined as $G_{\sigma \sigma^\prime}(\epsilon )\equiv
\langle\langle c_{\sigma}|\,c^+_{\sigma'}\rangle\rangle_\epsilon
$. Writing equation of motion for $\langle\langle
c_{\sigma}|\,c^+_{\sigma'}\rangle\rangle_\epsilon $, one arrives
at
\begin{eqnarray}
&&(\epsilon -\epsilon_{d})\langle\langle
c_{\sigma}|\,c^+_{\sigma'}\rangle\rangle_{\epsilon}
=\delta_{\sigma\sigma'}
\nonumber\\
&&+\sum_{k}[ T^{*l}_{k\beta}\langle\langle
a_{lk\beta}|\,c^+_{\sigma'}\rangle\rangle_{\epsilon}
+T^{*r}_{k\beta}\langle\langle
a_{rk\beta}|\,c^+_{\sigma'}\rangle\rangle_{\epsilon}\cos
(\varphi/2)
\nonumber\\
&&-{\beta}T^{*r}_{k-\beta}\langle\langle
a_{rk-\beta}|\,c^+_{\sigma'}\rangle\rangle_{\epsilon}\sin
(\varphi/2)]+U\langle\langle{c_{\sigma}n_{-\sigma}|\,c^+_{\sigma'}\rangle\rangle_{\epsilon}},
\nonumber\\
&&
\end{eqnarray}
where $\beta =+$ for $\sigma = \uparrow$ and $\beta =-$ for
$\sigma = \downarrow$. Applying equation of motion to the four new
Green functions on the r.h.s. of Eq.(7), one finds
\begin{equation}
(\epsilon-\varepsilon^l_{k\beta})\langle\langle
a_{lk\beta}|\,c^+_{\sigma'}\rangle\rangle_{\epsilon}=
T^{l}_{k\beta}\langle\langle
c_{\sigma}|\,c^+_{\sigma'}\rangle\rangle_\epsilon,
\end{equation}
\begin{eqnarray}
&&(\epsilon-\varepsilon^r_{k\beta})\langle\langle
a_{rk\beta}|\,c^+_{\sigma'}\rangle\rangle_{\epsilon}=
T^{r}_{k\beta}\langle\langle
c_{\sigma}|\,c^+_{\sigma'}\rangle\rangle_\epsilon\cos(\varphi/2)
\nonumber\\
&& +\beta T^{r}_{k\beta}\langle\langle
c_{-\sigma}|\,c^+_{\sigma'}\rangle\rangle_\epsilon\sin(\varphi/2),
\end{eqnarray}
\begin{eqnarray}
&&(\epsilon-\varepsilon^r_{k-\beta})\langle\langle
a_{rk-\beta}|\,c^+_{\sigma'}\rangle\rangle_{\epsilon}=
T^{r}_{k-\beta}\langle\langle
c_{-\sigma}|\,c^+_{\sigma'}\rangle\rangle_\epsilon\cos(\varphi/2)
\nonumber\\
&& -\beta T^{r}_{k-\beta}\langle\langle
c_{\sigma}|\,c^+_{\sigma'}\rangle\rangle_\epsilon\sin(\varphi/2),
\end{eqnarray}
\begin{widetext}
\begin{eqnarray}
&&(\epsilon -\epsilon_{d}-U)\langle\langle
c_{\sigma}n_{-\sigma}|\,c^+_{\sigma'}\rangle\rangle_{\epsilon}
=\langle\{c_{\sigma}n_{-\sigma},\,c^+_{\sigma'}\}\rangle +
\sum_{k}[ T^{*l}_{k\beta}\langle\langle
a_{lk\beta}|\,c^+_{\sigma'}\rangle\rangle_{\epsilon}
+T^{*r}_{k\beta}\langle\langle
a_{rk\beta}\,n_{-\sigma}|\,c^+_{\sigma'}\rangle\rangle_{\epsilon}\cos
(\varphi/2)
\nonumber\\
&&
-\beta T^{*r}_{k-\beta}\langle\langle
a_{rk-\beta}\,n_{-\sigma}|\,c^+_{\sigma'}\rangle\rangle_{\epsilon}\sin
(\varphi/2) -T^{l}_{k-\beta}\langle\langle c_\sigma
a^+_{lk-\beta}c_{-\sigma}|\,c^+_{\sigma'}\rangle\rangle_{\epsilon}-
T^{*l}_{k-\beta}\langle\langle c_\sigma
a_{lk-\beta}c^+_{-\sigma}|\,c^+_{\sigma'}\rangle\rangle_{\epsilon}
\nonumber\\
&&
 -T^{r}_{k-\beta}\langle\langle c_\sigma
a^+_{rk-\beta}c_{-\sigma}|\,c^+_{\sigma'}\rangle\rangle_{\epsilon}\cos
(\varphi/2)- \beta T^{r}_{k\beta}\langle\langle c_\sigma
a^+_{rk\beta}c_{-\sigma}|\,c^+_{\sigma'}\rangle\rangle_{\epsilon}\sin
(\varphi/2)
\nonumber\\
&&
 -T^{*r}_{k-\beta}\langle\langle c_\sigma a_{rk-\beta}
c^+_{-\sigma}|\,c^+_{\sigma'}\rangle\rangle_{\epsilon}\cos
(\varphi/2) -\beta T^{*r}_{k\beta}\langle\langle c_\sigma
a_{rk\beta}c^+_{-\sigma}|\,c^+_{\sigma'}\rangle\rangle_{\epsilon}\sin
(\varphi/2)].
\end{eqnarray}
\end{widetext}

Now, the Hartree-Fock decoupling scheme is applied to the
higher-order Green functions generated on the r.h.s. of eq. (11),
\begin{equation}
\langle\langle a_{\nu
k\pm\beta}\,n_{-\sigma}|\,c^+_{\sigma'}\rangle\rangle_\epsilon
\rightarrow \langle n_{-\sigma}\rangle\langle\langle a_{\nu
k\pm\beta}|c^+_{\sigma'}\rangle\rangle_\epsilon,
\end{equation}
\begin{equation}
\langle\langle c_\sigma a_{\nu
k\pm\beta}\,c^+_{-\sigma}|\,c^+_{\sigma'}\rangle\rangle
\rightarrow \langle c^+_{-\sigma}c_\sigma\rangle\langle\langle
a_{\nu k\pm\beta}|c^+_{\sigma'}\rangle\rangle,
\end{equation}
\begin{equation}
\langle\langle c_\sigma a^+_{\nu
k\pm\beta}\,c_{-\sigma}|\,c^+_{\sigma'}\rangle\rangle\simeq 0,
\end{equation}
which closes the set of equations (7)-(11) and allows to find
solution for the causual Green functions $G_{\sigma
\sigma^\prime}(\epsilon )$. Here, $\langle ...\rangle$ means the
quantum statistical average value of the appropriate operator.

The general solution for the Green function $G_{\sigma
\sigma^\prime}(\epsilon )$ in the Hartree-Fock approximation may
be written in the matrix form as
\begin{equation}
\textbf{G}(\epsilon )=[\textbf{1}-\textbf{g}(\epsilon
)\mathbf\Sigma^{(0)}(\epsilon )]^{-1}\textbf{g}(\epsilon )=
[\textbf{1}-\textbf{g}^{(0)}(\epsilon )\mathbf\Sigma(\epsilon
)]^{-1}\textbf{g}^{(0)}(\epsilon ),
\end{equation}
where
$g^{(0)}_{\sigma\sigma^\prime}(\epsilon
)=\delta_{\sigma\sigma^\prime}(\epsilon-\varepsilon_d)^{-1}$ is
the Green function of the dot in the absence of both Coulomb
interaction and coupling to the leads, whereas
$\textbf{g}(\epsilon)$ is the corresponding Green function of the
dot decoupled from the electrodes but with the Coulomb interaction
included,
\begin{equation}
g_{\sigma\sigma}(\epsilon)=\frac{\epsilon-\varepsilon_d-U(1-\langle
n_{-\sigma}\rangle
)}{(\epsilon-\varepsilon_d)(\epsilon-\varepsilon_d-U)},
\end{equation}
\begin{equation}
g_{\sigma-\sigma}(\epsilon)=-\frac{U\langle
n_{-\sigma\sigma}\rangle}{(\epsilon-\varepsilon_d)(\epsilon-\varepsilon_d-U)},
\end{equation}
with $\langle n_{-\sigma\sigma}\rangle =\langle
c^+_{-\sigma}c_{\sigma}\rangle$. In Eq.(15)
$\mathbf\Sigma^{(0)}(\epsilon )$ is the self-energy of the
noninteracting ($U=0$) quantum dot, whereas
$\mathbf\Sigma(\epsilon )$ is the full self-energy including
contributions from the coupling to the leads and from the Coulomb
interaction. The self-energy $\mathbf\Sigma^{(0)}(\epsilon )$
takes the form,
\begin{equation}
\mathbf\Sigma^{(0)}(\epsilon )=\left(
\begin{array}{cc}
\Sigma^{(0)}_{+}(\epsilon ) & \Sigma^{(0)}_1(\epsilon )  \\
\Sigma^{(0)}_1(\epsilon ) & \Sigma^{(0)}_{-}(\epsilon )   \\
\end{array}
\right),
\end{equation}
where
\begin{eqnarray}
&&\Sigma^{(0)}_{\pm} (\epsilon )=\sum_{k}
\frac{|T^{l}_{k\pm}|^2}{\epsilon -\varepsilon^l_{k\pm}}
\nonumber\\
&& +\sum_{k} \left( \frac{|T^{r}_{k\pm}|^2}{\epsilon
-\varepsilon^r_{k\pm}}\cos^{2}(\varphi/2)
 +\frac{|T^{r}_{k\mp}|^2}{\epsilon
-\varepsilon^r_{k\mp}}\sin^{2}(\varphi/2)] \right),
\nonumber\\
\end{eqnarray}
\begin{equation}
\Sigma^{(0)}_{1}(\epsilon )=\frac{1}{2}\sum_{k} \left(
\frac{|T^{r}_{k+}|^2}{\epsilon -\varepsilon^r_{k+}} -
\frac{|T^{r}_{k-}|^2}{\epsilon -\varepsilon^r_{k-}} \right) \sin
\varphi \,.
\end{equation}
As follows from Eq.(15), the full self-energy
$\mathbf\Sigma(\epsilon )$ is determined by
$\mathbf\Sigma^{(0)}(\epsilon )$ and ${\mathbf g}^{-1}(\epsilon )$
{\it via} the relation
\begin{equation}
\mathbf\Sigma(\epsilon )=\mathbf\Sigma^{(0)}(\epsilon )+{\mathbf
g}^{(0)-1}(\epsilon ) - {\mathbf g}^{-1}(\epsilon ).
\end{equation}
The retarded and advanced Green functions $\mathbf G^{R(A)}
(\epsilon)$ can be found as $\mathbf G^{R(A)} (\epsilon)=\mathbf
G(\epsilon\pm i\eta)$. Similarly, one can find
$\mathbf\Sigma^{R(A)} (\epsilon)$ and $\mathbf\Sigma^{(0)R(A)}
(\epsilon)$. From Eqs (18) to (20) follows that the retarded
self-energy $\mathbf\Sigma^{(0)\mathnormal{R}}(\epsilon)$ has the
general form (18) with the components given explicitly by the
formulae
\begin{eqnarray}
\Sigma^{(0)R}_{\pm}(\epsilon)&=&-\frac{1}{2}
\Gamma^l_{\pm}(\epsilon) \left[ \frac{1}{\pi}\ln\left(
\frac{D+eV_l-\epsilon}{D-eV_l+\epsilon} \right) +i \right]
\nonumber\\
&& -\frac{1}{2}\left[ \Gamma^r_{\pm}(\epsilon)\cos^2(\varphi/2)
+\Gamma^r_{\mp}(\epsilon)\sin^2(\varphi/2) \right]
\nonumber\\
&& \times\left[ \frac{1}{\pi}\ln\left(
\frac{D+eV_r-\epsilon}{D-eV_r+\epsilon} \right)+i \right] ,
\end{eqnarray}
\begin{eqnarray}
\Sigma^{(0)R}_{1}(\epsilon)&=&-\frac{1}{4}\left[
\Gamma^r_{+}(\epsilon)-\Gamma^r_{-}(\epsilon) \right]\sin \varphi
\nonumber\\
&&
 \times \left[ \frac{1}{\pi}\ln\left( \frac{D+eV_r
-\epsilon}{D-eV_r+\epsilon} \right)+i \right],
\end{eqnarray}
where
\begin{equation}
\Gamma^\nu_{\pm}(\epsilon)=2\pi\sum_{k} \vert T_{k\pm}^\nu\vert^2
\delta (\epsilon -\epsilon_{k\pm}^\nu )
\end{equation}
for $\nu =l,r$.  It has been assumed that the lower and upper
edges of the electron band at zero bias are at $-D$ and $D$,
respectively.

In the following we assume
\begin{equation}
\Gamma^l_{\pm}(\epsilon )=\Gamma^l_{\pm}=\Gamma_0(1\pm{p_{\,l}})
\end{equation}
and
\begin{equation}
\Gamma^r_{\pm}(\epsilon
)=\Gamma^r_{\pm}=\alpha\Gamma_0(1\pm{p_{\,r}})
\end{equation}
when $\epsilon$ is within the electron band and zero otherwise.
The parameters $p_{\,l}$ and $p_{\,r} $ describe the spin
asymmetry of the coupling to the left and right electrodes,
respectively, $\Gamma_0$ is a constant, and the parameter $\alpha$
takes into account asymmetry between coupling of the dot to the
left and right electrodes.

The key problem is to find the correlation Green function
$\textbf{G}^<(\epsilon)$. This can be calculated from the Keldysh
equation,
\begin{equation}
\textbf{G}^<(\epsilon)=\textbf{G}^R(\epsilon)\mathbf\Sigma^{\mathnormal{<}}
(\epsilon)\textbf{G}^{\mathnormal{A}}(\epsilon),
\end{equation}
provided the lesser self-energy
$\mathbf\Sigma^{\mathnormal{<}}(\epsilon)$ is known. However,
finding $\mathbf\Sigma^{\mathnormal{<}} (\epsilon)$ for an
interacting ($U>0$) system is a difficult tusk and one has to use
some approximate solutions. One approach is the so-called Ng
ansatz, according to which the lesser (greater) self energy
$\mathbf\Sigma^{\mathnormal{<(>)}} (\epsilon)$ is proportional to
the self-energy of the corresponding non-interacting system,
$\mathbf\Sigma^{\mathnormal{<(>)}} (\epsilon)=\mathbf
A\mathbf\Sigma^{(0)\mathnormal{<(>)}} (\epsilon)$, where $\mathbf
A$ can be be determined on taking into account the relation
$\mathbf\Sigma^{\mathnormal{<}} (\epsilon)
-\mathbf\Sigma^{\mathnormal{>}}
(\epsilon)=\mathbf\Sigma^{\mathnormal{R}}
(\epsilon)-\mathbf\Sigma^{\mathnormal{A}} (\epsilon)$. Taking into
account the above and Eq.(21) one finds $\mathbf A =\mathbf I$,
where $\mathbf I$ is the unit matrix. Thus, in the Hartree-Fock
approximation one finds
\begin{equation}
\mathbf\Sigma^{\mathnormal{<}}(\epsilon)=-\sum_\nu
[\mathbf\Sigma^{(0)\mathnormal{R}}_{\nu}(\epsilon)
-\mathbf\Sigma^{(0)\mathnormal{A}}_{\nu}(\epsilon)]{\mathnormal{f_\nu}}(\epsilon),
\end{equation}
where $f_\nu (\epsilon)$ is the Fermi-Dirac distribution function
for the $\nu$-th electrode, $f_\nu (\epsilon) =1/\{1+\exp[(
\epsilon-\mu_\nu )/k_BT]\}$, with the electrochemical potentials
$\mu_l=eV_l=eV/2$ and $\mu_r=eV_r=-eV/2$. (The energy is measured
from the Fermi level of the leads in equilibrium.) In the
Hartree-Fock approximation the relation (28) can also be obtained
from the equation of motion method.

The average values of the occupation numbers $\langle
n_{\sigma}\rangle =\langle c^+_\sigma c_\sigma\rangle $ and
$\langle n_{\sigma-\sigma}\rangle =\langle c^+_\sigma
c_{-\sigma}\rangle $, which enter the expressions for Green
functions, have to be calculated self-consistently by using the
formulae
\begin{equation}
\langle n_\sigma\rangle ={\rm
Im}\int^{+\infty}_{-\infty}\frac{d\epsilon}{2\pi}G^<_{\sigma\sigma}(\epsilon)
\end{equation}
and
\begin{equation}
\langle n_{\sigma-\sigma}\rangle
=-i\int^{+\infty}_{-\infty}\frac{d\epsilon}{2\pi}G^<_{ -\sigma
\sigma}(\epsilon).
\end{equation}

\section{Transport characteristics}

Having found the Green functions, one can calculate electric
current, spin accumulation, and spin precession on the dot. To do
this we calculate the average values of all the three spin
components, which are related to the diagonal and off-diagonal
occupation numbers (Eqs (35) and (36)). The corresponding
relations may be written as (spin components are measured in the
units of $\hbar$),
\begin{equation}
\langle S_z\rangle =(n_\uparrow -n_\downarrow )/2 ,
\end{equation}
\begin{equation}
\langle S_y\rangle ={\rm Im} (n_{\uparrow\downarrow}),
\end{equation}
\begin{equation}
\langle S_x\rangle = {\rm Re} (n_{\uparrow\downarrow}),
\end{equation}
where $n_\uparrow$, $n_\downarrow$, and $n_{\uparrow \downarrow}$
are calculated self-consistently, as described above.

In turn, electric current flowing from the $\nu$-th lead to the
dot is given by the formula \cite{jauho94}
\begin{equation}
J_\nu=\frac{ie}{\hbar}\int^{+\infty}_{-\infty}\frac{d\epsilon}{2\pi}{\rm
Tr}\left\{{\mathbf{\Gamma}}_\nu (\textbf{G}^<(\epsilon ) + f_\nu
(\epsilon )[\textbf{G}^R(\epsilon )-\textbf{G}^A(\epsilon
)])\right\},
\end{equation}
with
\begin{equation}
{\mathbf{\Gamma}}_l=\Gamma_0 \left(
\begin{array}{cc}
  1+p_{\,l} & 0 \\
  0 & 1-p_{\,l} \\
\end{array}
\right)
\end{equation}
and
\begin{equation}
{\mathbf{\Gamma}}_r=\Gamma_0\alpha \left(
\begin{array}{cc}
  1+p_{\,r}\cos\varphi  & p_{\,r}\sin\varphi \\
  p_{\,r}\sin\varphi &  1-p_{\,r}\cos\varphi \\
\end{array}
\right).
\end{equation}
Thus, taking into account Eqs (22)-(34), together with Eqs
(40)-(42), one obtains the final symmetrized expression for
electric current, $J=(1/2)(J_l-J_r)$, in the form
\begin{equation}
J=\frac{e\alpha\Gamma^2_0}{4\pi\hbar}\int^{+\infty}_{-\infty}d\epsilon
[f_l(\epsilon)-f_r(\epsilon)] j(\epsilon),
\end{equation}
where
\begin{widetext}
\begin{eqnarray}
j(\epsilon)&=&2(1+p_{\,l})(1+p_{\,r}\cos\varphi)G^R_{\uparrow\uparrow}(\epsilon)G^{R*}_{\uparrow\uparrow}(\epsilon)
+2(1-p_{\,l})(1-p_{\,r}\cos\varphi)G^R_{\downarrow\downarrow}(\epsilon)G^{R*}_{\downarrow\downarrow}(\epsilon)
\nonumber\\
 &&
+[(1+p_{\,l})(1-p_{\,r}\cos\varphi)+(1-p_{\,l})
(1+p_{\,r}\cos\varphi)]
[G^R_{\uparrow\downarrow}(\epsilon)G^{R*}_{\uparrow\downarrow}(\epsilon)+G^R_{\downarrow\uparrow}(\epsilon)G^{R*}_{\downarrow\uparrow}(\epsilon)]
\nonumber\\
 &&
+(1+p_{\,l})p_{\,r}\sin\varphi
\{G^R_{\uparrow\uparrow}(\epsilon)[G^{R*}_{\uparrow\downarrow}(\epsilon)+
G^{R*}_{\downarrow\uparrow}(\epsilon)]+G^{R*}_{\uparrow\uparrow}(\epsilon)[G^{R}_{\uparrow\downarrow}(\epsilon)+
G^{R}_{\downarrow\uparrow}(\epsilon)]\}
\nonumber\\
 &&
+(1-p_{\,l})p_{\,r}\sin\varphi
\{G^R_{\downarrow\downarrow}(\epsilon)[G^{R*}_{\uparrow\downarrow}(\epsilon)+
G^{R*}_{\downarrow\uparrow}(\epsilon)]+G^{R*}_{\downarrow\downarrow}(\epsilon)[G^{R}_{\uparrow\downarrow}(\epsilon)+
G^{R}_{\downarrow\uparrow}(\epsilon)]\}.
\end{eqnarray}
\end{widetext}

The current formula derived above can be applied to any magnetic
configuration of the system, and thus can be used to determine
TMR. Generally, TMR is described quantitatively by the ratio
\begin{equation}
TMR=\frac{R(\varphi)-R(\varphi =0)}{R(\varphi
=0)}=\frac{J(\varphi=0)-J(\varphi)}{J(\varphi)},
\end{equation}
where $J(\varphi)$ is the electric current flowing through the
system when the angle between spin polarizations of the leads is
$\varphi$, and $R(\varphi)$ is the corresponding electrical
resistance.

\section{Numerical results}

\subsection{Symmetrical junctions}

In a symmetrical case both barriers are identical, $\alpha=1$, and
the electrodes are made of the same ferromagnetic material,
$p_l=p_r$. Consider first electronic transport in a symmetrical
junction with fully polarized (half-metallic) external electrodes,
$p_{\,l}=p_{\,r}=1$, and with the dot level above the Fermi level
of the electrodes at equilibrium, $ \varepsilon_d>0$.

Bias dependence of electric current and the corresponding
differential conductance are shown in Fig.2(a) and Fig.2(b) for
selected values of the angle $\varphi$. The current-voltage curve
for parallel configuration ($\varphi=0$) reveals typical step-like
characteristics. Below the first (lower) threshold voltage the dot
is empty and thus sequential contribution to electric current is
exponentially suppressed. The first step in the current occurs at
the bias, where the discrete level $\varepsilon_d$ crosses the
Fermi level of the source electrode (the dot can be occupied by a
single electron), whereas the step at a higher voltage (higher
threshold) corresponds to the case when $\varepsilon_d+U$ crosses
this Fermi level (the dot may be doubly occupied).

In non-collinear configurations a monotonous suppression of the
tunneling current with increasing $\varphi$ takes place in the
whole bias voltage range, and the current disappears for $\varphi
=\pi$. This is a typical (perfect) spin-valve effect. Suppression
of electric current is due to an electron residing on the dot,
whose spin orientation prevents it from tunneling to the drain
lead. In the extreme case of antiparallel configuration,
$\varphi=\pi$, the electron that has tunneled to the dot from the
fully polarized source electrode blocks transport through the
junction since it cannot tunnel further to the oppositely
polarized drain lead. This scenario holds as long as spin-flip
relaxation processes are absent.

The steps in current-voltage characteristics give rise to the
narrow peaks in differential conductance displayed in Fig.2(b),
which occur at the lower and higher threshold voltages. Apart from
this, dependence of electric current on magnetic configuration of
the system leads to the TMR effect, defined quantitatively by
Eq.(39), and is shown in Fig.2(c). The effect increases with
increasing angle $\varphi$ and tends to infinity when $\varphi
\rightarrow\pi$ (therefore there is no curve in Fig.2(c) for
$\varphi =\pi$).
\begin{figure}[h]
\begin{center}
\includegraphics[width=1\columnwidth,angle=0]{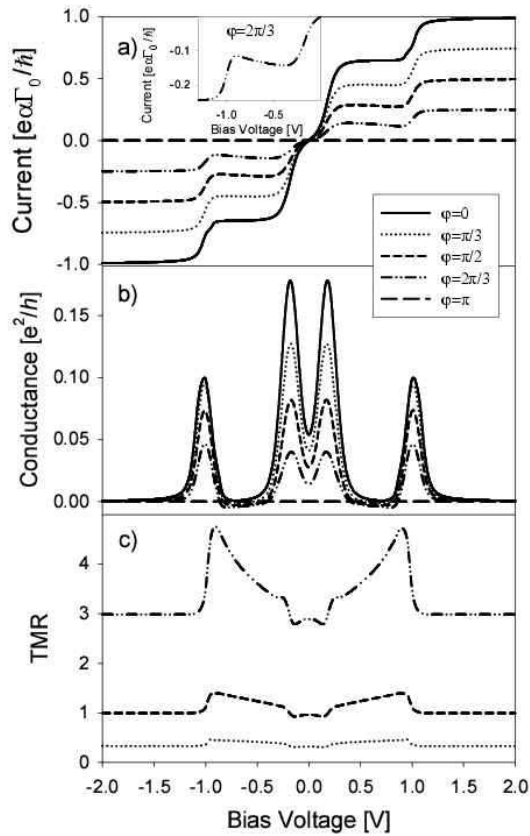}
\caption{\label{currentplots}Bias dependence of electric current
(a), differential conductance (b), and tunnel magnetoresistance
(c), calculated for indicated values of the angle $\varphi$. The
inset in (a) shows electric current between the lower and higher
threshold voltages for $\varphi=2\pi/3$. The parameters assumed
for numerical calculations are: $\varepsilon_d=0.1$ eV, $U=0.4$
eV, $\Gamma_0 = 0.01$ eV, $p_l=p_r=1$, $\alpha =1$ and $T=100$ K.}
\end{center}
\end{figure}
\begin{figure}[h]
\begin{center}
\includegraphics[width=1\columnwidth,angle=0]{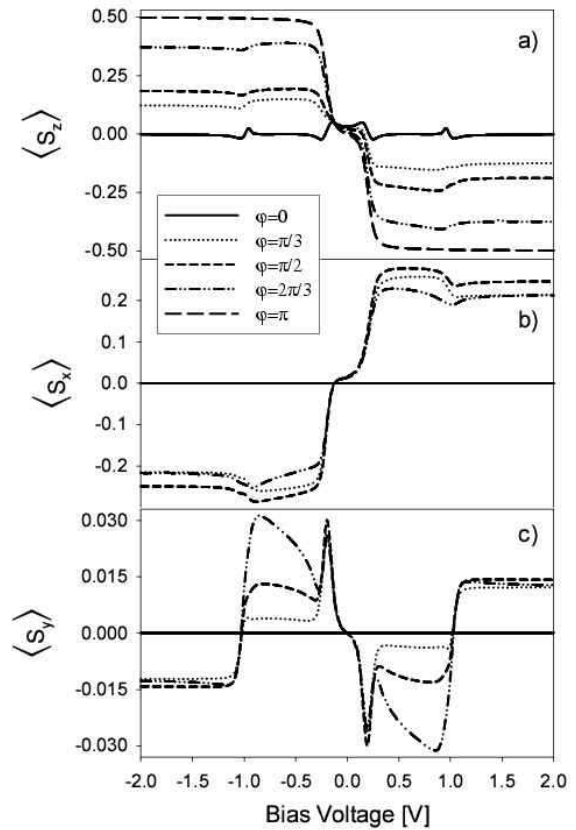}
\caption{\label{currentplots}Bias dependence of the average spin
components; $\langle S_z\rangle$ (a), $\langle S_x\rangle$ (b) and
$\langle S_y\rangle$ (c) for indicated values of the angle
$\varphi$. The other parameters are as in Fig.2.}
\end{center}
\end{figure}

An interesting feature of the current-voltage characteristics is
the negative differential conductance, which may occur in
non-collinear configurations between the lower and higher
threshold voltages (between the corresponding two peaks in the
differential conductance). The negative differential conductance
corresponds to some enhancement in TMR, as can be seen in
Fig.2(c). The enhancement is particularly significant for rather
large values of the angle $\varphi$ (smaller that $\pi$). Physical
origin of this feature becomes clear, when taking into account
bias dependence of the average spin on the dot in nonequilibrium
situation (spin accumulated on the dot).
\begin{figure}[h]
\begin{center}
\includegraphics[width=1\columnwidth,angle=0]{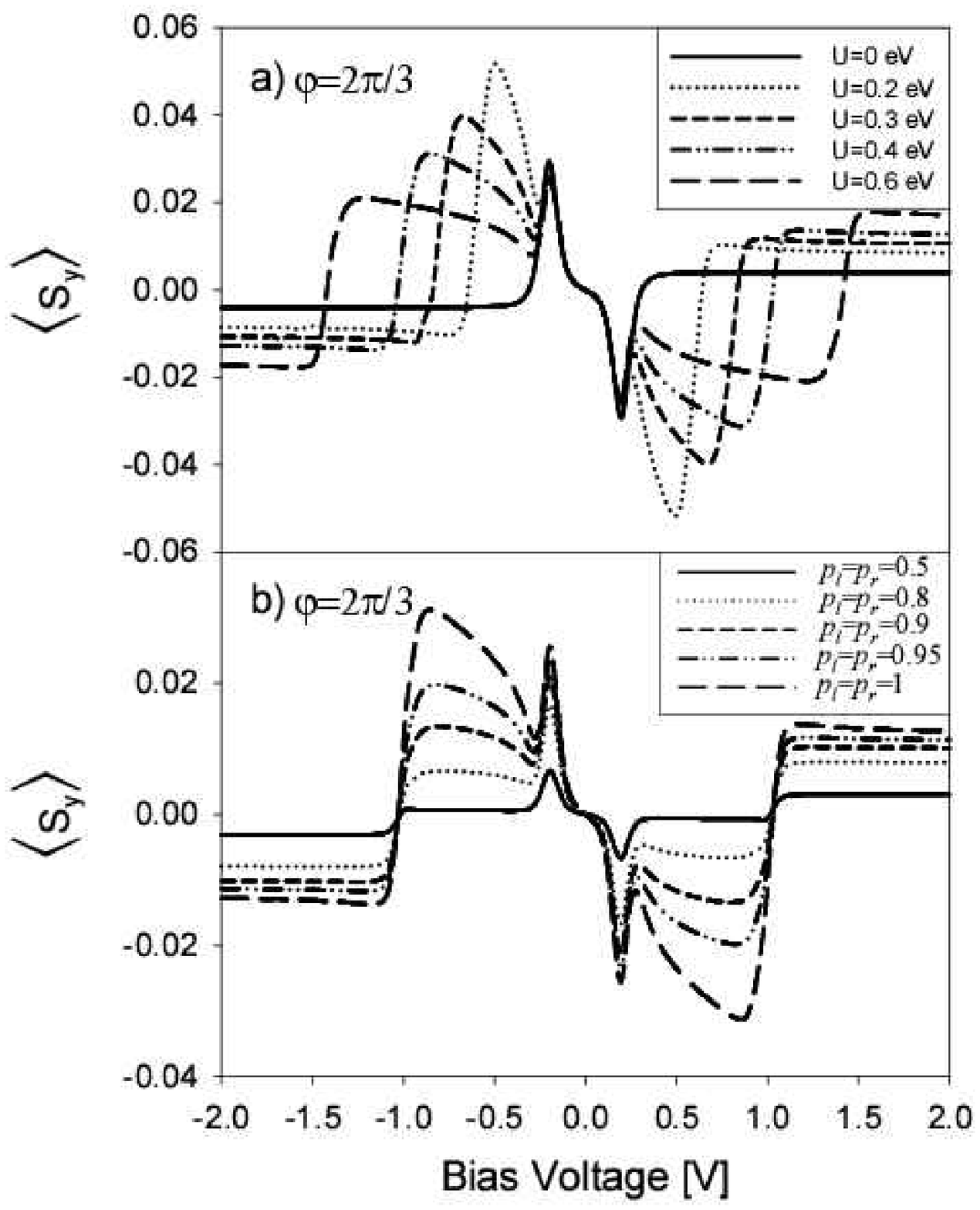}
\caption{\label{currentplots} Bias dependence of the average spin
component $\langle S_y\rangle$ for $\varphi=2\pi/3$ and for
indicated values of the Coulomb correlation parameter $U$ (a) and
the lead polarization (b). The other parameters are as in Fig.2.}
\end{center}
\end{figure}
As shown in Fig.3(a), the absolute value of the average $\langle
S_z\rangle$ increases with increasing $\varphi$ and almost
vanishes in the parallel configuration. In turn, the average value
of the $x$-component, $\langle S_x\rangle$, vanishes for both
parallel and antiparallel alignments and is nonzero for canted
configurations, as shown in Fig.3(b). Similarly, the average value
of $\langle S_y\rangle$ also vanishes in the collinear
configurations and is nonzero in the noncollinear ones. One should
point here, that neither initial spin state in the source
electrode, nor the final state in the drain electrode have
nonvanishing $y$-component (perpendicular to the plane determined
by spin polarizations of the two leads). What is then the reason
of nonvanishing $\langle S_y\rangle$? This can be accounted for by
taking into account the fact, that an electron residing on the dot
experiences a certain exchange field resulting from coupling
between the dot and leads, which effectively acts as a local
magnetic field \cite{braun04}. Strength of this molecular field
and its orientation with respect to the global quantization axis
depend on the applied bias voltage and on the angle between the
magnetic moments of the leads. Thus, if an electron that has
tunneled from the source lead resides sufficiently long time on
the dot level, its spin experiences a torque due to the exchange
interaction, which results in precession of the average spin
around the molecular field, and consequently in a nonzero average
value of the transverse component $\langle S_y\rangle$. As shown
in Fig.3(c), this precession-induced component is significant in
non-collinear cases (see the case of $\varphi=2\pi/3$) and in the
voltage range between the two threshold voltages. Just this
enhanced precession is associated with a decrease in electric
current, which leads to negative differential conductance.

As follows from the above discussion, the magnitude of $\langle
S_y\rangle$ is a measure of the spin precession induced by the
effective exchange field. In Fig.4(a) we show $\langle S_y\rangle$
for different values of the Coulomb correlation parameter $U$. The
curve for $U=0$ indicates that spin precession in noncollinear
configurations takes place also when there is no Coulomb
interaction between electrons on the dot. However, as follows from
Fig.4(a), the presence of such an interaction enhances the spin
precession and also extends the voltage range where the precession
is significant.

\begin{figure}[h]
\begin{center}
\includegraphics[width=1\columnwidth,angle=0]{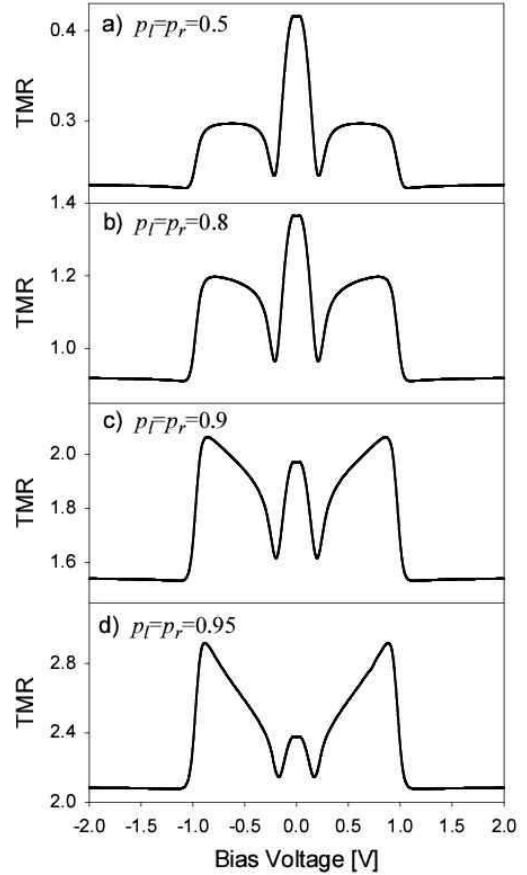}
\caption{\label{currentplots}Bias dependence of the tunnel
magnetoresistance in symmetrical junctions for indicated values of
the lead polarization. The curves are plotted for the
non-collinear configuration, $\varphi=2\pi/3$. The other
parameters are as in Fig.2.}
\end{center}
\end{figure}
\begin{figure}[h]
\begin{center}
\includegraphics[width=1\columnwidth,angle=0]{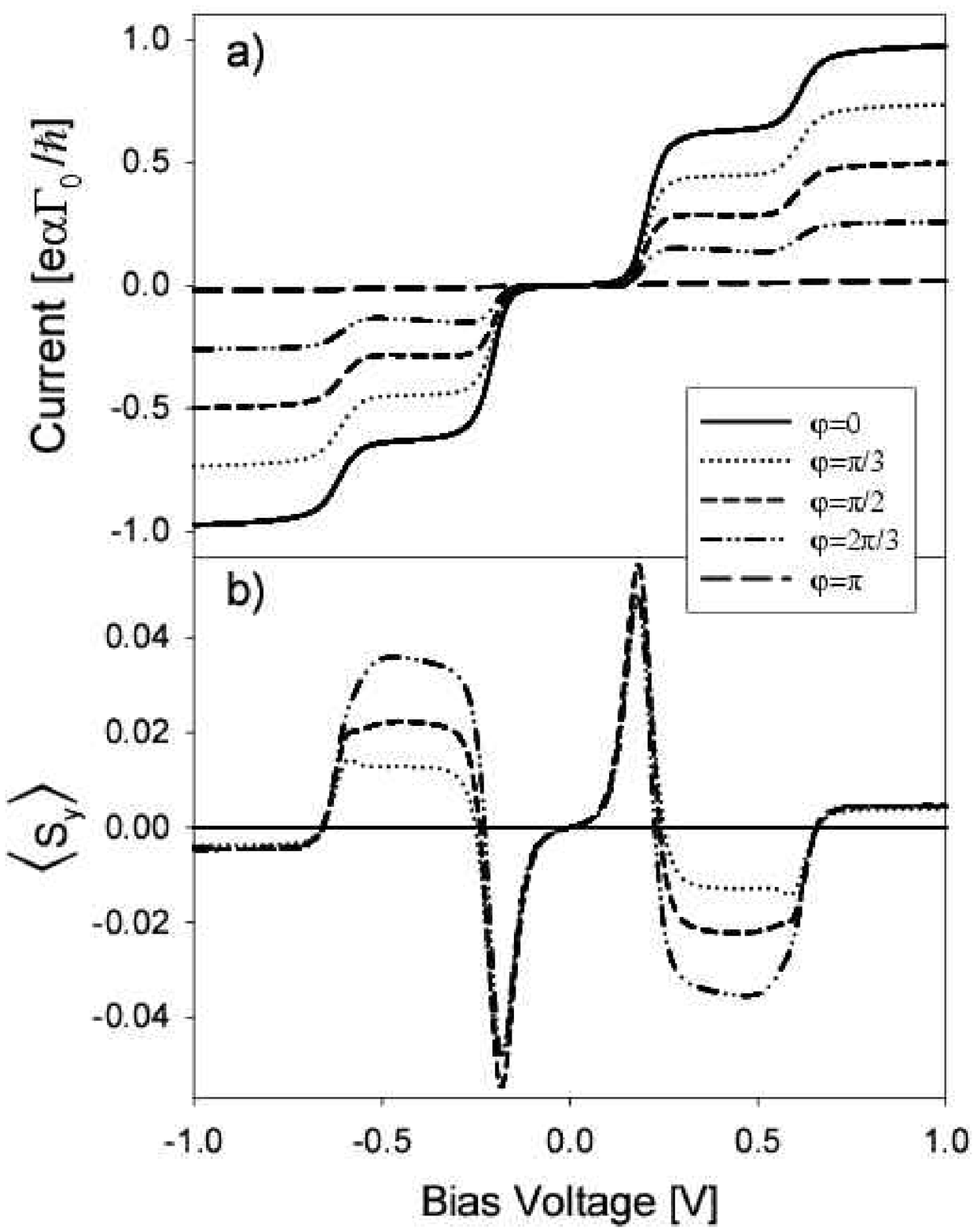}
\caption{\label{currentplots}Bias dependence of electric current
(a), and the average value of the $y$ component of the dot spin
(b), calculated for indicated values of the angle $\varphi$ and
$\varepsilon_d=-0.1$ eV. The other parameters are as in Fig.2.}
\end{center}
\end{figure}
The spin precession also depends on the spin polarization of
external electrodes. This is shown in Fig.4(b), where the
$y$-component of the average spin accumulated on the dot is shown
for several values of the parameters $p_l$ and $p_r$. This figure
clearly shows that the spin precession decreases when spin
polarization of the leads becomes smaller. This is reasonable
since lower spin polarization creates smaller exchange field.

When the spin polarization rate  of the leads becomes smaller than
1, electric current can flow also in the antiparallel
configuration, contrary to the case shown in Fig.2 for
$p_l=p_r=1$. Consequently, the TMR effect also becomes smaller and
remains finite for $\varphi =\pi$. In Fig.5 we show the TMR effect
for different polarizations of the external electrodes, and for a
particular noncollinear configuration ($\varphi=2\pi/3$). The
curves are symmetrical with respect to the bias reversal. The
central peak corresponds to an enhanced magnetoresistance in the
current blockade regime. Such an enhancement of TMR in the
blockade regime, where sequential tunneling is suppressed, was
also observed earlier for collinear configurations
\cite{swirkowicz02}. Although the sequential current is
exponentially suppressed in the blockade regime, the electrons
still can flow due to higher order processes (like cotunneling).
The other two broad maxima placed symmetrically on both sides of
the central peak, occur between the two (lower and higher)
threshold voltages. The following features of these two maxima are
interesting to note. First, for small values of the polarization
parameters, the central maximum is larger than the others. The
situation changes with increasing polarization factors, and now
the central peak becomes smaller for high spin polarizations of
the leads. Second, the two maxima become strongly asymmetric for
large values of the spin polarization factors, as clearly visible
in Fig.5. To account for this behavior one should take into
account bias dependence of $\langle S_y\rangle$ from Fig.4(b),
which shows clearly the corelations between the spin precession
and height and shape of the TMR peaks with increasing
polarization.

In the situation studied above the dot level was empty at
equilibrium. Qualitatively similar behavior of electric current,
magnetoresistance, and average spin on the dot, has been found for
the situation when the dot level is below the Fermi level, and the
dot is occupied by a single electron at equilibrium. Exemplary
current-voltage characteristics are shown in Fig.6(a), and the
corresponding $y$-components of the average spin on the dot are
\begin{figure}[h]
\begin{center}
\includegraphics[width=1\columnwidth,angle=0]{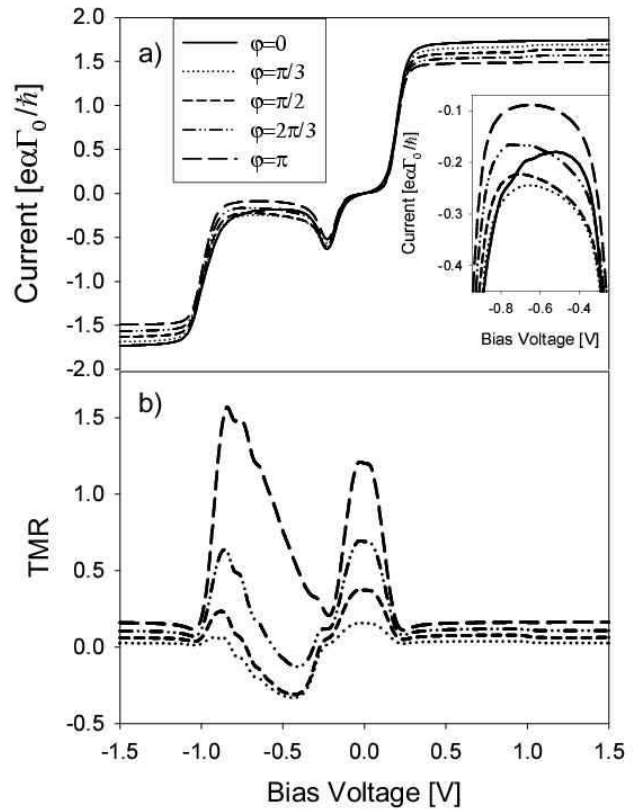}
\caption{\label{currentplots}Bias dependence of electric current
(a) and magnetoresistance (b), calculated for indicated values of
the angle $\varphi$. The inset in (a) shows current between the
threshold voltages for negative bias. The parameters assumed for
numerical calculations are: $\varepsilon_d=0.1$ eV, $U=0.4$ eV,
$\Gamma_0 = 0.01$ eV, $p_l=0.4$, $p_r=1$, $\alpha =0.1$ and
$T=100$ K.}
\end{center}
\end{figure}
\begin{figure}[h]
\begin{center}
\includegraphics[width=1\columnwidth,angle=0]{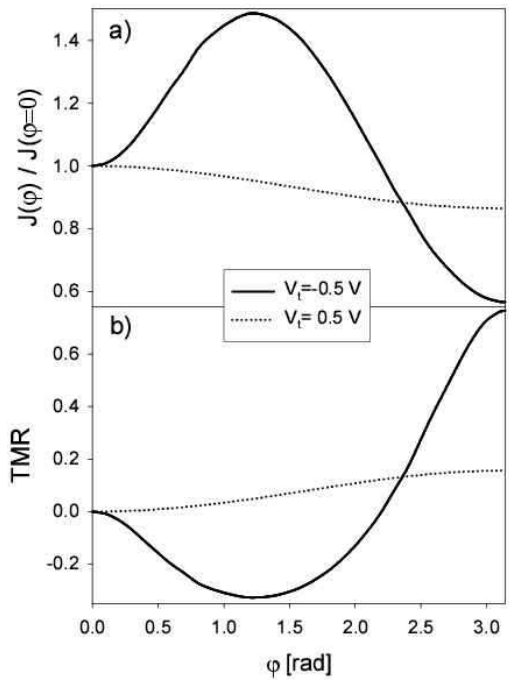}
\caption{\label{currentplots}Angular variation of electric current
(a) and TMR (b) for indicated values of the bias voltage. The
other parameters are as in Fig.7.}
\end{center}
\end{figure}
\begin{figure}[h]
\begin{center}
\includegraphics[width=1\columnwidth,angle=0]{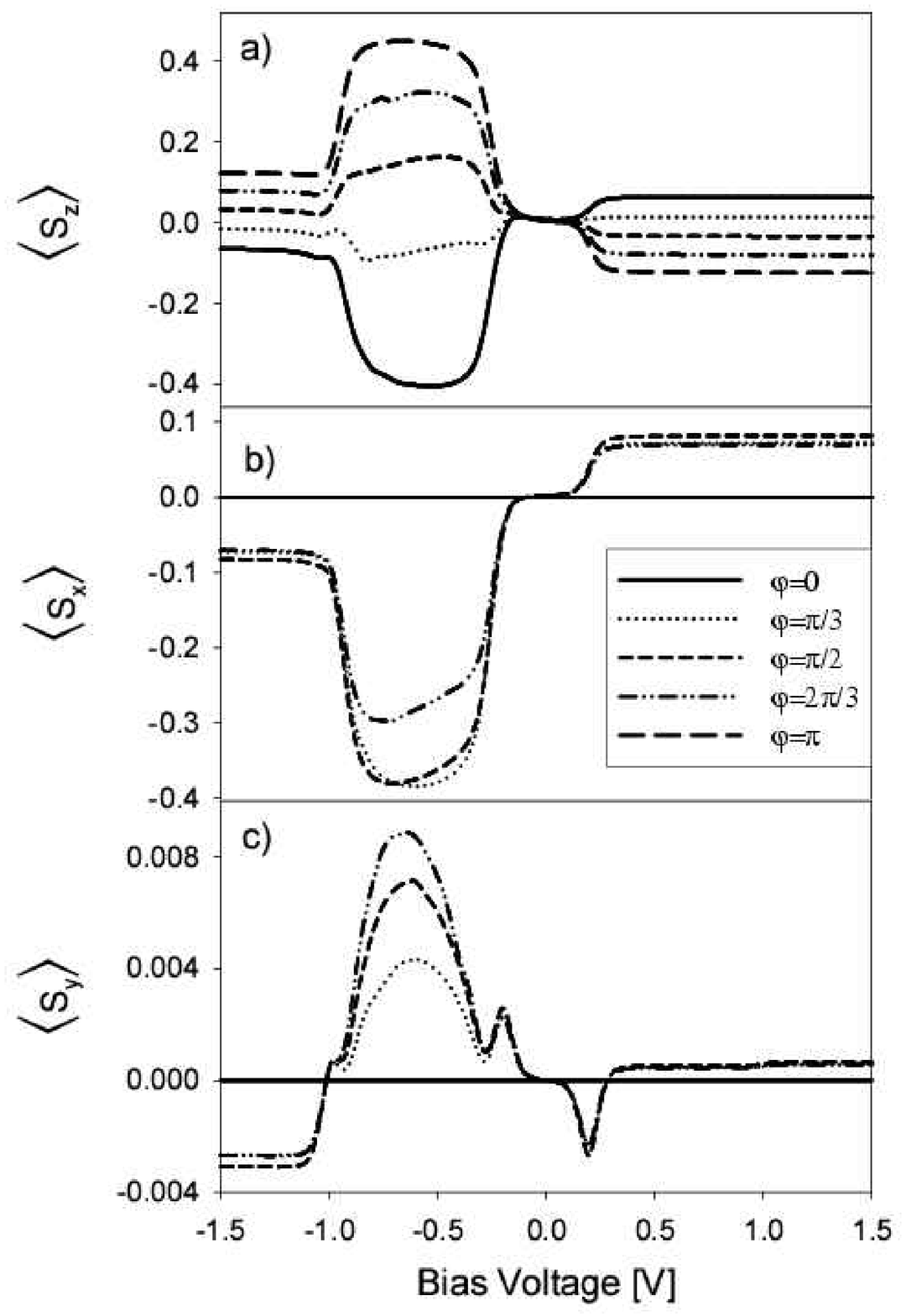}
\caption{\label{currentplots}Bias dependence of the average spin
components; $\langle S_z\rangle$ (a), $\langle S_x\rangle$ (b) and
$\langle S_y\rangle$ (c) for indicated values of the angle
$\varphi$. The other parameters are as in Fig.6.}
\end{center}
\end{figure}
\begin{figure}[h]
\begin{center}
\includegraphics[width=1\columnwidth,angle=0]{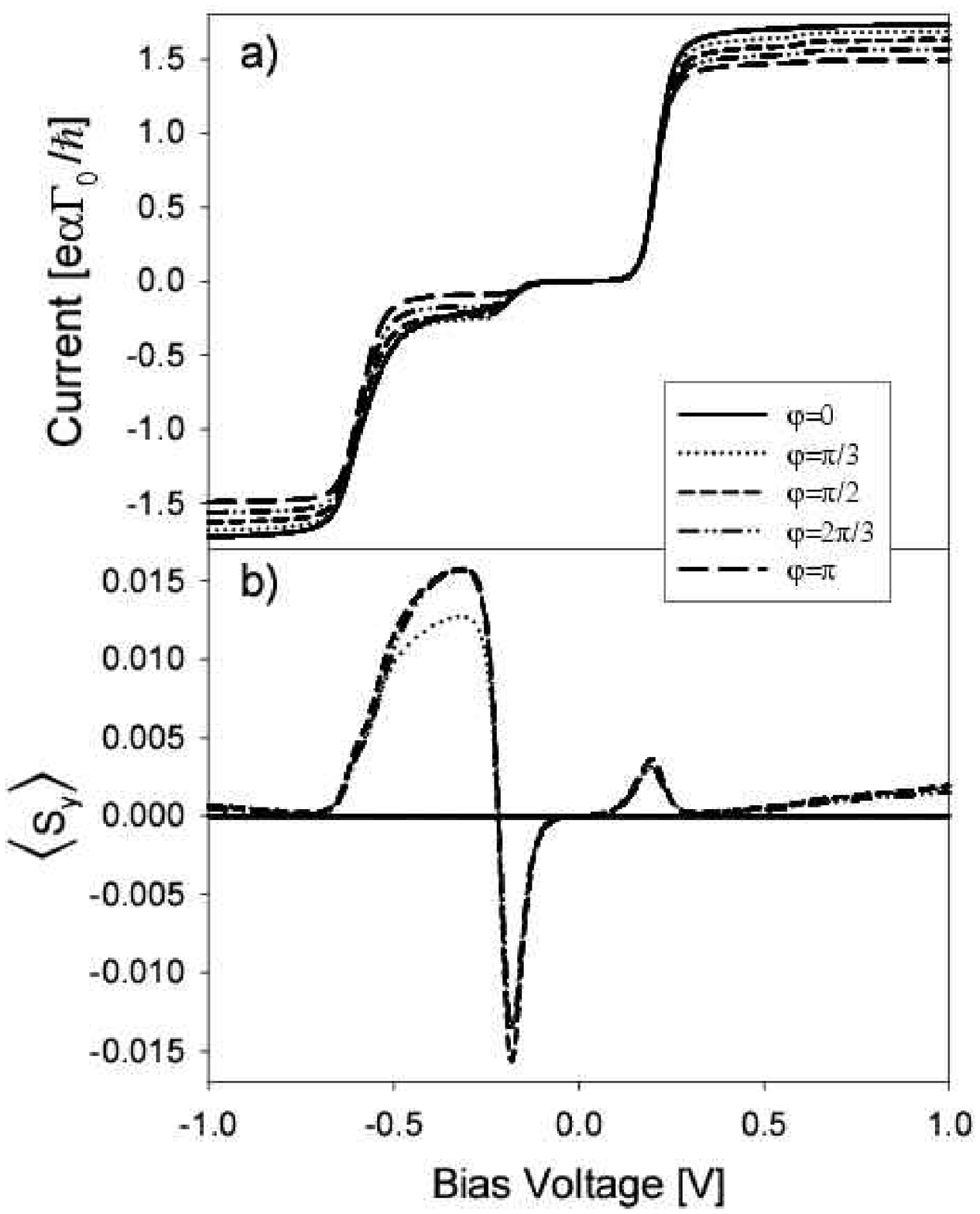}
\caption{\label{currentplots}Bias dependence of electric current
(a), and the average value of the $y$ component of the dot spin
(b), calculated for indicated values of the angle $\varphi$ and
$\varepsilon_d=-0.1$ eV. The other parameters are as in Fig.7.}
\end{center}
\end{figure}
shown in Fig.6(b). The current-voltage curves display similar
features as the curves shown in Fig.2(a), with characteristic
negative differential conductance between the threshold voltages
for noncollinear configurations. The main difference in the bias
dependence of $\langle S_y\rangle$ is a reversed sign of the peaks
at the lower threshold voltage in comparison to that in Fig.3(c).

\subsection{Asymmetrical junctions}

Now the dot is separated from both electrodes by non-equivalent
barriers, $\alpha\ne 1$, and  the electrodes are made of different
ferromagnetic materials, $p_l\ne p_r$. For numerical calculations
we assume $p_{\,l}$=0.4, $p_{\,r}$=1, and $\alpha =0.1$. More
specifically, it is assumed that the right electrode is made of a
half-metallic material with electrons being totally spin-polarized
at the Fermi level, whereas the factor $\alpha =0.1$ indicates
that on average electrons can tunnel much easier to (from) the
left electrode than to (from) the right one. This asymmetry
between the left and right electrodes and barriers gives rise to
asymmetrical transport characteristics of the junction with
respect to the bias reversal. The asymmetry is clearly visible in
current-voltage characteristics and bias dependence of TMR, shown
in Fig.7(a) and Fig.7(b), respectively. For positive bias (the
right lead is the source electrode), the current and TMR curves
are rather uniform above the first threshold voltage. The current
flows there for arbitrary value of the angle $\varphi$ and thus
TMR is significantly suppressed. The situation changes
diametrically when the electric current flows in the opposite
direction, i.e. when electrons tunnel through the dot from the
left electrode to the right (half-metallic) one. Below the first
threshold voltage sequential tunneling is exponentially suppressed
and only the higher-order tunneling processes are possible. When
the energy level $\varepsilon_d$ enters the tunneling window,
electric current starts to flow through the junction but this
takes place only in a small voltage range in the vicinity of the
first threshold voltage, where the characteristic resonant bump is
observed. Above the bump, the current is suppressed by an electron
residing on the dot. When $\varepsilon_d+U$ crosses the Fermi
level of the source lead, the current increases again and finally
saturates at a certain level. For positive bias, $V>0$, the curves
in Fig.7(a) and Fig.7(b) for different values of the angle
$\varphi$ reflect a monotonous angular variation of the current
and TMR, shown also explicitly in Fig.8 for a particular value of
the bias voltage. For negative bias one observes a more complex
and interesting behavior of the transport characteristics. First,
the above mentioned suppression of electric current between the
two steps is now less pronounced, and the corresponding angular
variation of electric current and TMR is non-monotonous, as shown
explicitly in Fig.8 by the relevant curve. When the negative bias
voltage surpasses the second threshold, the monotonous variation
of electric current and TMR is restored. It is also interesting to
note, that the TMR effect is enhanced in the voltage regions,
where electric current is suppressed. This is the region below the
first threshold voltage, and for negative bias also the region
between the two threshold voltages. The latter one is particularly
interesting as the TMR may change there sign from positive to
negative (see Fig.7(b) and Fig.8(b)).

Suppression of electric current by an electron of a given spin
orientation localized on the dot can be accounted for by analyzing
spin accumulated on the dot when a steady state current flows
through the system. This is illustrated in Fig.9(a-c), where the
part (c) shows the component induced by spin precession. The spin
precession is particularly enhanced for negative bias between the
two thresholds - exactly where electric current is suppressed. The
enhancement is a consequence of relatively long time that
electrons spend on the dot.

The situation is different for the dot level lying under the Fermi
level, i.e., when the dot is already occupied by one electron in
equilibrium situation. The corresponding numerical results are
shown in Fig.10, where part (a) shows the current-voltage
characteristics, and part (b) the average value of the
perpendicular $y$-component of the average spin accumulated on the
dot. The transport characteristics for collinear configurations
were already accounted for in earlier publications
\cite{rudzinski01}, and their most interesting feature is the
pronounced asymmetry with respect to the bias reversal (diode-like
behavior). For noncollinear configurations transport
characteristics have new features, which are qualitatively similar
to those found in the case of empty dot at equilibrium, like for
instance nonmonotonous angular variation of electric current and
TMR.

\section{Summary and conclusions}

Using the non-equilibrium Green function approach we have
calculated electric current, average value of electron spin
accumulated on the dot, and tunnel magnetoresistance due to
rotation of the magnetic moments of external electrodes. It has
been shown that the average spin precesses by a certain angle
around an effective exchange field arising from the interaction
between the dot and electrodes. This precession leads to a nonzero
value of $\langle S_y\rangle$, i.e., of the component normal to
the plane determined by magnetic moments of both electrodes.

It has been also shown that the spin precession is enhanced by
Coulomb correlations and strong spin polarization of the leads.
The spin precession is shown to exist also in the limit of
vanishing Coulomb correlations on the dot. Moreover, the interplay
of Coulomb correlations and effective exchange field may lead to a
negative differential conductance in the voltage range between the
two threshold voltages. It has been also shown, that the
diode-like  features of the  system  are partially suppressed when
magnetic  moments of the electrodes become non-collinear.

\begin{acknowledgments}
The work was supported by the State Committee for Scientific
Research through the Research Project PBZ/KBN/044/P03/2001 and 4
T11F 014 24. One of us (JB) also acknowledges useful
correspondence with Karsten Flensberg.
\end{acknowledgments}

\end{document}